# Long baseline neutrino beams at Fermilab

**S Childress and J Strait**
Fermilab, Batavia, IL 60510, U.S.A.
E-mail: childress@fnal.gov, strait@fnal.gov

**Abstract.** Fermilab has had a very active long baseline neutrino program since initiation of the NuMI project in 1998. Commissioned in 2005, the NuMI beam with 400 kW design power has been in operation for the MINOS neutrino oscillation program since that time. Upgrade of NuMI to 700 kW for NOvA is now well advanced, with implementation of the beam upgrades to be accomplished in 2012-2013. Design development for the next generation LBNE neutrino beam is now a major ongoing effort. We report here salient features and constraints for each of these beams, as well as significant challenges both experienced and expected.

## 1. Introduction

The Fermilab long-baseline neutrino beams discussed here are of conventional design, but with far greater beam power than for historical neutrino beams. Accelerated protons strike a target to produce charged pions, which are magnetically focused, then decay to neutrinos in a following drift region. A hadron absorber is at the end of the decay region, backed by appropriate muon shielding prior to a near detector on the Fermilab site. Neutrinos then continue on a direction between the near detector and a far detector at a distance of hundreds of kilometres. To achieve the needed far detector neutrino event rates we must have very intense proton beams, challenging target and focusing systems, plus massive neutrino detectors. These very powerful conventional design neutrino beams at megawatt levels have become known as 'Super beams'.

**Table 1.** Fermilab long baseline neutrino beams

|  | NuMI/MINOS | NuMI/NOvA | LBNE |
|---|---|---|---|
| Proton beam power | 0.4 MW | 0.7 MW | 0.7 to 2.3 MW |
| Proton energy | 120 GeV | 120 GeV | 60 to 120 GeV |
| Repetition rate | 1.87 sec (design) | 1.33 sec | 1.33 sec |
| Protons per spill | $4.0 \times 10^{13}$ | $4.9 \times 10^{13}$ | $4.9 \times 10^{13}$ to $1.6 \times 10^{14}$ |
| Baseline | 735 km | 810 km | 1300 km |

## 2. NuMI – Neutrinos at the Main Injector

The 400 kW NuMI beam was designed to enable precision measurement of muon neutrino oscillations by the MINOS experiment. Beam design parameters were determined before oscillation parameters were known, with tunability of the neutrino beam spectrum a key requirement.

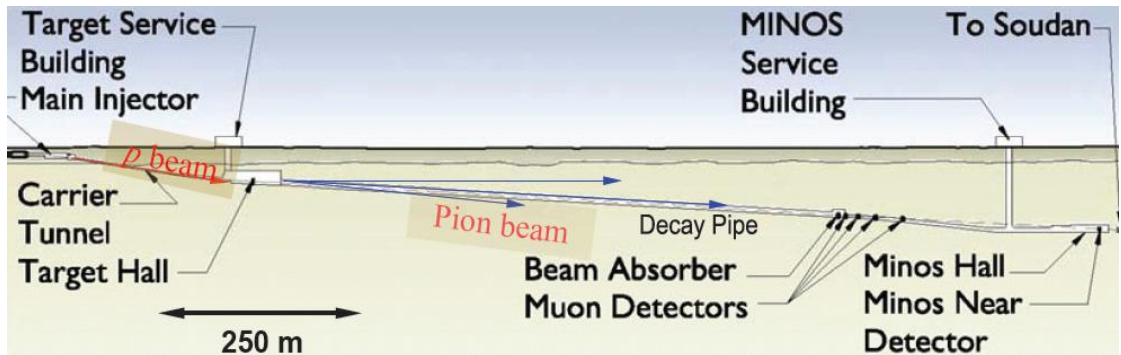

**Figure 1.** Profile view of NuMI beam layout.

A schematic view of the NuMI target, focusing horns and decay region is shown in figure 2. The 675 meter long decay allowed the possibility of a high energy neutrino beam when the horn separation was also increased by moving the second horn (reflector) downstream. With better knowledge of oscillation parameters prior to NuMI operation, the chosen operational configuration shown in the figure was for lower neutrino energy. Within this configuration a more limited range of neutrino energies peaked from 3 to 9 GeV is still feasible by moving the target longitudinally relative to the first horn, with the target significantly inside the horn for the normal low energy operation. This range of neutrino energies is shown in figure 3.

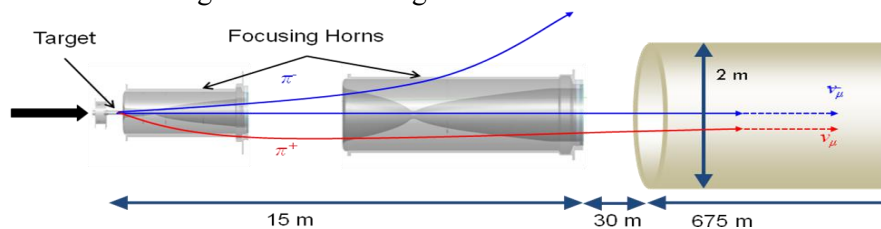

**Figure 2.** Focusing horns in neutrino mode collect positive pions while removing negatives. The resultant muon neutrino beam is 91.7% pure. An antineutrino beam is created by reversing the horn polarity

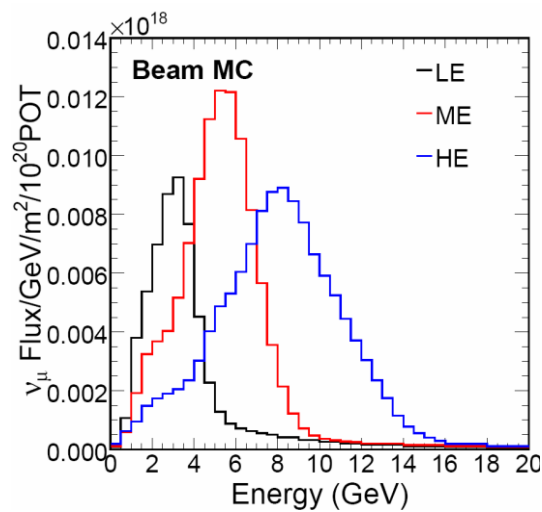

**Figure 3.** Range of neutrino energies achievable by moving target into (LE) or away from the first horn (ME, HE). Normal MINOS operation is in LE position, with average neutrino energy of ~ 3 GeV. ME and HE target position data is used for tuning beam flux models.

*2.1 NuMI beam operation*

The NuMI beam has been in operation for six years, with a considerable amount of experience gained during this process. Over this time, average beam power during operation has been ~ 250 kW, with a continuing increase in power. Recent operation has been at ~ 330 kW, with shorter periods near the design peak of 400 kW. A total of $1.3 \times 10^{21}$ protons on target at 120 GeV have been accumulated with more than 50 million extracted beam pulses. Discussed below is our experience for some key NuMI systems.

*2.2 Proton beam*

A significant early concern with intense NuMI beam operation was for radiological protection of an aquifer region through which the unshielded primary beam tunnel passes prior to targeting in a deep target hall mined in the underlying dolomite rock. Another consideration was the need for rigorous protection against significant activation of the multi-ton magnets positioned on steep 100 to 150 milliradian slopes. Additionally, beam technical components can be damaged within a few seconds by an off trajectory proton beam.

Many improvements beyond previous beam operation were developed to address these issues. These include a comprehensive beam permit system which monitors status of more than 250 inputs prior to enabling each extracted NuMI beam pulse, open extraction and primary transport geometries to accept a range of extracted beam conditions, fully automated beam position control with no manual control of beam positions during operation, and high precision power supply regulation. In the tunnel radiation safety monitoring requires fractional beam loss for each extraction to be $< 10^{-5}$ with normal fractional loss at a few $\times 10^{-7}$ of the transported beam. All of this has worked extremely well, and going forward provides capability for precision control of megawatt proton beams.

*2.3 Target hall systems*

NuMI target hall components are subjected to very severe conditions, with resultant challenging learning experiences for both targets and focusing horns during our high intensity beam operation. Early failures with both systems prior to the readiness of spares necessitated the development of efficient repair capabilities for the highly activated components. These capabilities have enabled overall efficient NuMI beam operation over a period of years in spite of many challenges.

Two failure modes have been experienced with the focusing horns. Original design electrical isolators for each horn water line had a Kovar interface connecting ceramic and stainless steel. These interfaces were prone to failure after millions of horn pulsings, with several failures during beam operation. New design replacement isolators use a pressure fit seal directly between ceramic and stainless steel, which has been problem free. A second type of horn system failure was experienced at the strip-line power feed, where high-strength steel washers failed due to hydrogen embrittlement. High-strength steel is no longer used for applications inside the target chase.

Experience with NuMI targets is shown in figure 4.

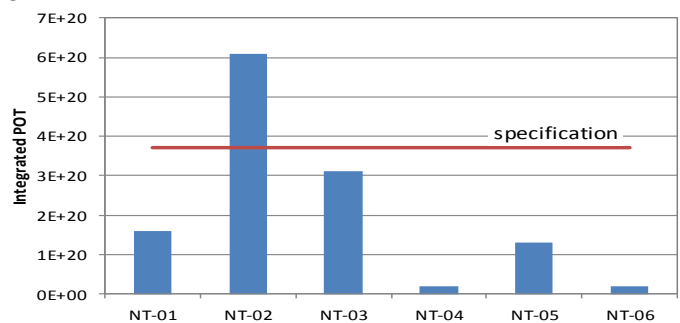

**Figure 4.** Integrated protons on target for the six NuMI graphite targets used 2005-2011. The horizontal line indicates the target design lifetime specification of $3.7 \times 10^{20}$ POT.

The second target NT-02 was problem free, and used for three years of NuMI beam operation. In later stages a progressive reduction in neutrino flux to 15% was seen, which we attribute to graphite degradation from the $6 \times 10^{20}$ POT. The other 5 NuMI targets experienced failures related to cooling water leaks, with the worst experience for the last 3 targets produced. An extensive effort has evaluated the target weaknesses seen, with a number of improvements implemented in new production targets. Early operating experience with target NT-07 is positive.

Shown in figure 5 is a drawing of the target inserted into the first horn when in the normal low energy target position. Very tight spatial constraints exist with need for a low mass target system to minimize absorption of the low energy pions important for the desired neutrino spectrum.

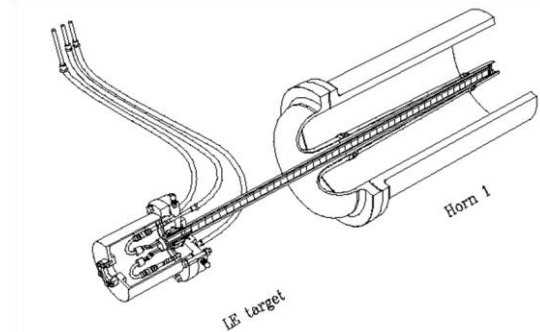

**Figure 5.** NuMI target inserted into horn when in normal low energy position

*2.4 Decay region*
The 675 meter NuMI decay volume was designed as a vacuum system with a thin upstream aluminum window to minimize pion absorption. After corrosion was discovered on the window at the location impacted by protons not interacting in the target, a change was made to fill the decay region with helium. This reduces the neutrino flux by a few percent, but greatly lessens pressure gradient across the window.

*2.5 Tritium mitigation*
Along with target reliability, this has been the most significant issue for sustained NuMI beam operation. The deep NuMI beam enclosure system was designed to let some ground water enter the tunnel complex over its full length. While initial inlet flow rates were higher, the current rate of 500 liters/minute is still substantial. This water is then pumped to the surface, and into the laboratory cooling ponds system. Designing for water inflow provided a significant cost advantage for the facility construction, and also insured that water near the enclosure perimeter would be collected and released as surface water where allowable tritium levels are significantly higher than in the underlying aquifer.

NuMI produces a few hundred Curies per year of tritium in hadron showers, which is initially imbedded in the radiation shielding. But significant amounts of this tritium can become highly mobile in the presence of moist air. For the NuMI beam, a dominant source of tritium in water pumped from the tunnel was determined to have been produced in shielding surrounding the target hall chase, with the hadron absorber shield providing a smaller source. Humid air passing through these shields, in time came in contact with tunnel water outside the shields.

A significant dehumidification system has been added to intercept the majority of tritium prior to precipitation into NuMI sump waters. This system has been augmented as beam power has increased, with additional mitigation efforts being evaluated. An additional significant advantage of reduced humidity in the target hall chase is less corrosion for target hall components.

## 3. NOvA Neutrino Experiment

Upgrade of the Fermilab accelerator complex and NuMI to 700 kW beam power is currently ongoing for the NOvA neutrino experiment. In parallel with the beam efforts, a new 15 kTon detector designed for precision measurement of oscillations by muon neutrinos to electron neutrinos is being constructed in northern Minnesota just south of the U.S.-Canada border. Initial NOvA beam is expected in 2013.

The NOvA far detector is off axis from the center trajectory of the NuMI beam by 14 milliradians, generating a narrower neutrino energy spectrum peaked at 2 GeV, as shown in figure 6.

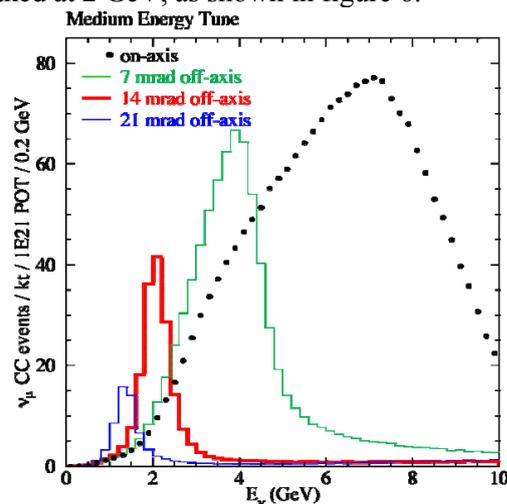

**Figure 6.** Effect on neutrino energy spectrum for progressively larger off axis angles. The 14 mrad spectrum peaked at 2 GeV is ideal for oscillation sensitivity at the NOvA detector baseline length of 810 km.

Accelerator improvements for NOvA take advantage of the 8 GeV Recycler storage ring, now available after completion of the Tevatron collider program. Twelve Booster accelerator batches will be injected into the Recycler, and then slip stacked into six batches for single turn extraction to the Main Injector accelerator. As this injection process occurs in parallel with the previous MI acceleration cycle, a major reduction in the cycle time between 120 GeV proton extractions to NOvA is possible. Most of the beam power increase is due to this reduction to 1.33 seconds cycle time. An additional intensity increase to $4.9\times10^{13}$ protons per cycle is also feasible, giving 700 kW beam power to NOvA at 120 GeV.

NuMI primary beam upgrades include power system improvements to enable the 1.33 second cycle, improved precision for major dipole regulation, and more robust magnets for higher current quadrupoles. For the target hall, a significantly more robust target design is made feasible with the target no longer constrained to fit inside the first horn. Horn design is improved to reduce beam heating effects, target chase cooling is augmented, and the reflector horn is positioned further downstream enabling the optimal neutrino spectrum at the off axis NOvA detector. The on axis MINOS far detector will now see a higher energy spectrum as is also shown in figure 6. The redesigned NOvA target is shown in figure 7.

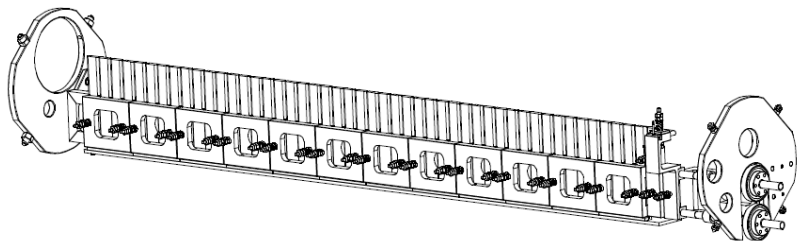

**Figure 7.** NOvA target. Beam passes through the graphite fins in upper part of the target canister.

## 4. LBNE – Long Baseline Neutrino Experiment

Physics goals for the next U.S long baseline neutrino oscillation experiment LBNE require a new beamline with capabilities not attainable with the existing NuMI/NOvA beam:

- Longer baseline to separate CP and matter effects, with 1300 km an optimal length.
- Beam power capability upgradeable to >2 MW when available with Project X beam.
- Beam optimized for lower energy and smaller electron neutrino component, obtainable with a shorter, wider decay pipe.
- Broad band beam covering $1^{st}$ and $2^{nd}$ oscillation maxima (2.5 and 0.8 GeV), with minimal high energy tail above ~ 5 GeV.

LBNE beam is designed to aim toward the Homestake mine in South Dakota, the site chosen by a national search process for the optimal far detector location. Beam trajectory is shown in figure 8.

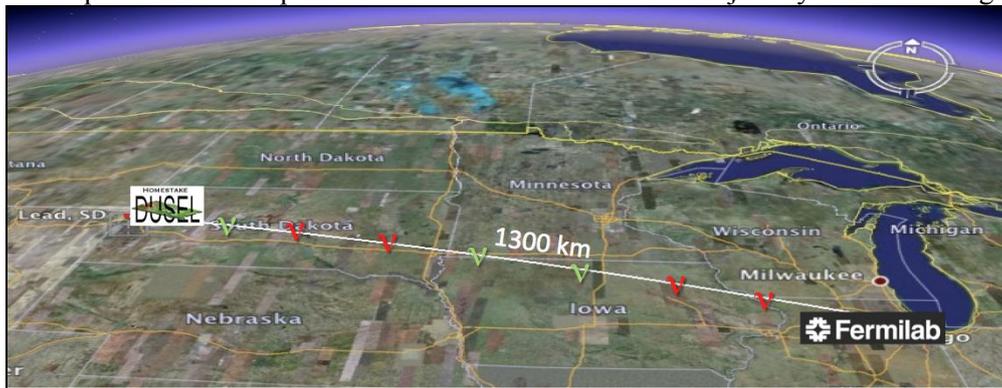

**Figure 8.** LBNE beam trajectory.

Other high level requirements for the LBNE beam are that initial beam power will be 700 kW available prior to Project X, with > 2 MW capability established for critical systems not readily upgradeable. Proton beam energy should be from 60–120 GeV. Beam design must meet stringent radiation safety requirements, and the LBNE design also should meet needed technical requirements while minimizing beam and facility cost.

We have evaluated two options for proton beam extraction from the Main Injector accelerator, the MI-60 northward extraction currently used for NuMI/NOvA and a westward MI-10 extraction requiring minimal additional bending for the trajectory toward South Dakota. Either option has good technical feasibility for the extraction design. MI-10 extraction is significantly more cost effective, but has more limited available space on the laboratory site for the full beam facility.

We have also evaluated two fundamentally different beam facility placements with either a deep mined beam facility as for NuMI, or with proton beam targeting above natural grade level. These two options are shown in figures 9 and 10.

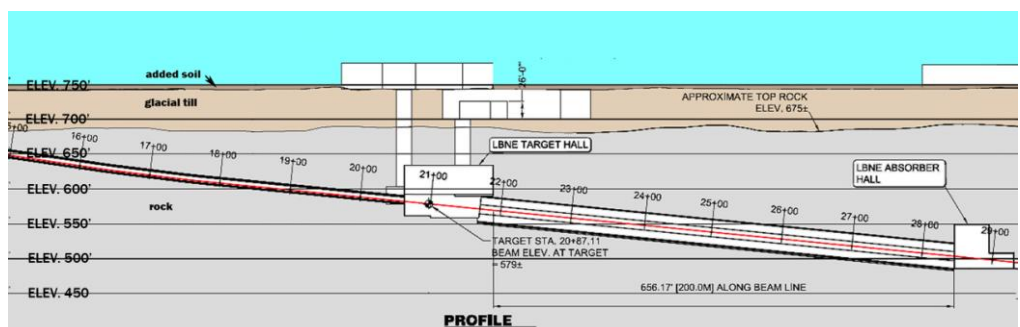

**Figure 9.** Profile view of a deep LBNE beam facility target hall and decay region.

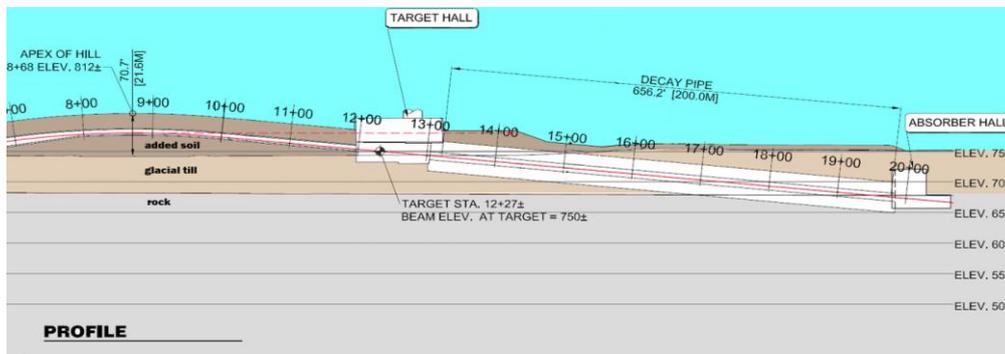

**Figure 10**. Profile view of a partially above grade LBNE beam facility target hall and decay region.

In each figure is shown the glacial till compacted clay region which extends ~ 20 meters below grade level, and then the underlying dolomite rock. For the above grade facility added soil above natural grade level is also shown.

For the deep facility option, placement of the target hall is chosen by reaching sufficient depth to enable a structural rock cover above the hall equal to the span of the hall. The facility is then mined in the supporting rock, as was done for NuMI construction. Required length of the proton beamline to reach this targeting depth precludes a 120 GeV proton beam with MI-10 extraction due to the limited site space, but is feasible for 60-90 GeV beams. With MI-60 beam extraction, ample site space is available. A possible advantage of the deep beam facility is that many features such as shielding and handling techniques are more readily understood due to similarity with NuMI.

For the partially above grade facility option, target hall placement is chosen to balance construction requirements for the elevated proton beam versus decay region and absorber hall located partially in rock. The only deep mined construction needed now at the Fermilab site is for the near detector hall (not shown here). The significantly shorter proton beamline now enables full energy range of 60-120 GeV with MI-10 extraction. To maintain needed component position stability, a key requirement with the above grade beam is for footing supports to underlying rock for the elevated proton beam enclosure and upstream decay region. A number of features for this beam facility design are quite different from previous site construction in the glacial till or from recent NuMI experience. Hence, there is the added requirement of a more challenging design learning curve. However, this beam facility option has two possibly compelling advantages, significantly reduced cost and major improvement for tritium mitigation.

*4.1 LBNE beam technical challenges*
Initial LBNE beam power of 700 kW is the same as we will have for NOvA beam operation, although some technical challenges as discussed below will be significant. Key features of LBNE beam designs are the need to enable the capability for a significant beam power upgrade when available with Project X and also to build beam systems and facility which can function with high efficiency for decades of operation. Discussed below are some challenges for key LBNE systems.

*4.2 Proton beam.*
Although many features for the LBNE proton beam are very different for both the two beam extraction options and also for the facility design options, the differences are ones of detail and are not fundamental. Technical solutions should be of similar difficulty regardless of selection choices, as also challenges of beam operation and control for ultimate LBNE beam power. Systems developed for NuMI, with iterative refinements underway for NOvA can be expected to work well for either LBNE proton beam.

As an example, for a deep LBNE proton beam ground water protection against loss of the very intense beam is an obvious challenge, while personnel shielding is simple with the large enclosure overburden. For an above grade beam ground water protection is easy, while significant attention is needed to insure appropriate personnel shielding. Component residual activation protection requirements are severe but similar for either design. In either case, rigorous control of the intense proton beam for every beam extraction, transport and targeting enables readily achievable solutions.

*4.2 Target hall systems*
The need for a relatively broad band low energy neutrino beam leads to design of an on-axis beam with target inside the first horn as for NuMI. Hence, we have the challenges of the NuMI target design, but for much higher beam power. This design challenge is well appreciated, and efforts are ongoing on several fronts including target material beam tests and design coordination with international experts. Experience gained with the NuMI target modifications and upcoming NOvA operation, as well as LBNE 700 kW operation should be productive in understanding good solutions for > 2MW LBNE beam.

A lesson from NuMI experience which is being incorporated into LBNE target hall design is the need for significantly improved remote handling capability with activated components, as well as improved time turn around for component repair or replacement. At the highest beam power, there will be many challenges with technical component design for target hall components and instrumentation in highly activated areas. Again, experience from 700 kW beam operation should aid significantly in understanding design upgrades which may be needed.

*4.4 Tritium mitigation*
Based on NuMI experience, the most important design features for tritium mitigation are to minimize humidity inside the target chase, decay and absorber shield, and to avoid mixing air release from these spaces with ground water sources. Addressing these issues in the LBNE facility design process is essential.

**Acknowledgments**
The authors wish to acknowledge inputs and assistance from many colleagues working with NuMI, NOvA and LBNE as well as the Fermilab accelerator systems

Fermilab is operated by Fermi Research Alliance, LLC under Contract No. DE-AC02-07CH11359 with the United States Department of Energy.